\documentclass[pre,twocolumn,aps,superscriptaddress,showpacs,floatfix]{revtex4}
\usepackage{amsmath}
\usepackage{amssymb}
\usepackage{amsbsy}
\usepackage{graphicx}
\usepackage{color}
 
\usepackage{enumerate}
\usepackage{mathtools}

\def\be{\begin{equation}}
\def\ee{\end{equation}}
\def\bfi{\begin{figure}}
\def\efi{\end{figure}}
\def\bea{\begin{eqnarray}}
\def\eea{\end{eqnarray}}

\def\norm{{\mathcal N}}

\begin{document}


\title{Heat fluctuations of Brownian oscillators in nonstationary processes: \\
fluctuation theorem and condensation transition}

\author{A. Crisanti}
\email{andrea.crisanti@uniroma1.it}
\affiliation{Dipartimento di Fisica, Universit\`a di Roma Sapienza, P.le Aldo Moro 2, 00185, Rome, Italy}
\affiliation{Istituto dei Sistemi Complessi - CNR, P.le Aldo Moro 2, 00185, Rome, Italy}

\author{A. Sarracino}
\email{alessandro.sarracino@roma1.infn.it}
\affiliation{Istituto dei Sistemi Complessi - CNR, P.le Aldo Moro 2, 00185, Rome, Italy}
\affiliation{Dipartimento di Fisica, Universit\`a di Roma Sapienza, P.le Aldo Moro 2, 00185, Rome, Italy}

\author{M. Zannetti}
\email{mrc.zannetti@gmail.com}
\affiliation{Dipartimento di Fisica ``E. R. Caianiello'', 
Universit\`a di Salerno, via Giovanni Paolo II 132, 84084 Fisciano (SA), Italy,}

\begin{abstract}
We study analytically the probability distribution of the heat
released by an ensemble of harmonic oscillators to the thermal bath,
in the nonequilibrium relaxation process following a temperature
quench. We focus on the asymmetry properties of the heat distribution
in the nonstationary dynamics, in order to study the forms taken by
the Fluctuation Theorem as the number of degrees of freedom is varied.
After analysing in great detail the cases of one and two oscillators,
we consider the limit of a large number of oscillators, where the
behavior of fluctuations is enriched by a condensation transition with
a nontrivial phase diagram, characterized by reentrant behavior.
Numerical simulations confirm our analytical findings. We also discuss
and highlight how concepts borrowed from the study of fluctuations in
equilibrium under symmetry breaking conditions [Gaspard,
  J. Stat. Mech. P08021 (2012)] turn out to be quite useful in
understanding the deviations from the standard Fluctuation Theorem.

\end{abstract}

\pacs{05.40.-a,05.70.Ln}

\maketitle

\section{Introduction}

At the level of the thermodynamical description of out of equilibrium
transformations, the irreversibility of macroscopic processes is
encoded in the second law expressed through strict inequalities
satisfied by various thermodynamic observables, like entropy and free
energies. At the more refined level of the statistical mechanical
description of the same processes, the irreversibility ought to
manifest as an asymmetry in the probability distributions of the
fluctuations of these same quantities.  The central and yet unsolved
problem is to find general methods, comparable to those available in
equilibrium, to characterize these probability distributions. A major
advance in this direction has been made in the last two decades with
the development of Fluctuations Theorems (FT), which allow to
constrain the form of the distributions with a certain degree of
universality and in conditions arbitrarily far from
equilibrium~\cite{Seifert}. This theoretical approach turned out to be
very useful, both in experimental and numerical studies, in the
characterization of several nonequilibrium systems, such as colloidal
particles in harmonic traps~\cite{Wang,Zon,imparato07,kim}, vibrated
granular media~\cite{Sarracino,Naert,Gnoli}, models of coupled
Langevin equations~\cite{Farago,Joubaud,Crisanti}, driven stochastic
Lorentz gases~\cite{Gradenigo,Gradenigo2}, and active
matter~\cite{Kumar}, just to name a few examples.

An important development in the field has been achieved with the understanding
of the impact of symmetry and symmetry breaking on the FT of currents
in stationary diffusive systems~\cite{Hurtado}.
More recently, it has been shown that relations in the form of FT are of
general occurrence also in the probability distributions of order
parameters in equilibrium statistical mechanics, in connection with
symmetry breaking~\cite{Gaspard}. Here we show that these equilibrium
results do feedback in the understanding of the nonequilibrium FT,
shedding new light on the conditions under which the FT in the
standard form (namely, with a linear asymmetry function) is expected to
hold~\cite{Puglisi,Baiesi,Harris,Park}. In particular, we study the
fluctuations of the heat exchanged with the environment by one or more
Brownian oscillators in an interval of time $[t_w,t]$, during the
relaxation following the instantaneous quench from high to low
temperature. This protocol induces a \emph{nonstationary dynamics}
that is noninvariant under time translation and therefore the heat
exchanged along a given trajectory explicitly depends on the two times
$(t_w,t)$. The study of the FT in similar processes has been addressed
in~\cite{Ritort,Picco,Zamponi}. Here, we derive exact expressions for
the heat probability density functions, for one, two and a large
number of independent oscillators. Our analysis shows how, in the case
of more than one oscillator, the heat probability obeys an FT which
deviates from the standard form and whose physical meaning can be
rationalized by resorting to the analogy with the equilibrium problem
in the presence of a nonuniform external field.

In the case of a large number of degrees of freedom, arising, for
instance, in the normal mode decomposition of an extended system, we
present a computation based on the steepest descent method, which
allows us to obtain an accurate description of the large deviation
function of the exchanged heat. Here, we will use
  large deviation theory with a large number of degrees of freedom
  (and not for long time intervals) \cite{hugo,ld}. This kind of
problem was considered previously in Refs.~\cite{Gonnella,Jo,Pechino},
showing that fluctuations undergo a condensation transition. Briefly,
by this is meant that fluctuations in a multi-component system do
condense if there exists a critical threshold above which the
fluctuation is feeded by {\it just one} of the components (or degrees
of freedom). Here, we analyse in detail the crossover region between
the normal and the condensed phase, providing an explicit expression
for the large deviation function. We also reconstruct numerically the
phase diagram at finite temperature, in the parameter space
$(t_w,\tau=t-t_w)$, which shows a nontrivial reentrant behavior.

The paper is organized as follows. In Section~\ref{zero} we present
the general framework within which various FT forms are derived and we
argue on physical grounds for the particular form we adopt.  We also
recall in some detail how an FT arises in statics as a consequence of
symmetry breaking, highlighting its relevance for the dynamical
problem.  Section~\ref{One} is devoted to the simplest case of a
single oscillator. We compute exactly the probability distribution of
heat fluctuations, we introduce the basic concept of time-dependent
effective temperature and we derive the FT in the ``Gallavotti-Cohen''
form~\cite{Gallavotti}. In Section~\ref{twobrownian} we consider the
case of two oscillators, and we analyse the modifications arising in
the FT form due to the presence of more than one degree of freedom.
In Section~\ref{extended} we consider the case of a large number of
oscillators. We discuss the conditions leading to condensation and we
map out numerically the phase diagram in the parameter space. The
consequences on the FT are analysed.  Finally, conclusions are drawn
in Section~\ref{conclusions}.

\section{General setup}
\label{zero}

There exist many variants of FT whose derivation~\cite{Maes,Seifert,Gawedzki}
can be unified into a single master theorem. Let us first outline the general setting
which applies to all fluctuation problems, in or out of equilibrium. 
Consider a sample space $\Omega$ with elements $\sigma$.
Let $\mu(\sigma)$ and $\mu^{\prime}(\sigma)$ be two arbitrary probability measures
over $\Omega$ and let ${\cal W}(\sigma)$ be defined by
\be
\frac{\mu^{\prime}(\sigma^*)}{\mu(\sigma)} = e^{-{\cal W}(\sigma)},
\label{I.1}
\ee
where the $*$ operation denotes an involutory transformation of $\Omega$ onto itself, 
i.e. $(\sigma^*)^*=\sigma$.
In the following this will be taken as the representation of the inversion element of
the $\mathbb{Z}_2 =(\mathbb{I},*)$ group, where $\mathbb{I}$ is the identity operator.
Introducing an arbitrary function ${\cal F}(\sigma)$, from the above relation there follows
the identity
\be
\langle {\cal F}(\sigma)e^{-{\cal W}(\sigma)} \rangle = \langle {\cal F}(\sigma^*) \rangle^{\prime},
\label{I.2}
\ee
where $\langle \cdot \rangle$ and $\langle \cdot \rangle^{\prime}$ denote expectations with respect
to $\mu(\sigma)$ and $\mu^{\prime}(\sigma)$, respectively. Taking for ${\cal F}(\sigma)$ the characteristic 
function $\theta_{{\cal M}}(\sigma|M)$ of a certain observable ${\cal M}(\sigma)$, that is
\be
\theta_{{\cal M}}(\sigma|M)  = \left \{ \begin{array}{ll}
        1,\;\; $if$ \;\; {\cal M}(\sigma) = M,\\
        0,\;\; $if$ \;\; {\cal M}(\sigma) \neq M,
        \end{array}
        \right .
        \label{I.2bis}
        \ee
from Eq.~(\ref{I.2}) follows
\be
\frac{P^{\prime}({\cal M}(\sigma^*) = M)}{P({\cal M}(\sigma) = M)} = 
\langle e^{-{\cal W}(\sigma)} | {\cal M}(\sigma) = M \rangle,
\label{I.3}
\ee
where $P^{\prime}$ and $P$ are the probabilities of the events in the arguments induced
by $\mu^{\prime}(\sigma)$ and $\mu(\sigma)$, respectively, and
in the right hand side there appears the expectation with respect to $\mu(\sigma)$,
conditioned to ${\cal M}(\sigma) = M$. Defining the $*$ transformation on the
set of random variables by ${\cal M}(\sigma) \mapsto {\cal M}^*(\sigma) = {\cal M}(\sigma^*)$,
which is also an involution, the above relation can be
recast in the form
\be
\frac{P^{\prime}({\cal M}^* = M)}{P({\cal M} = M)} = 
e^{-{\cal K}(M)},
\label{I.03}
\ee
with
\be
{\cal K}(M) = -\ln \langle e^{-{\cal W}(\sigma)} | {\cal M}(\sigma) = M \rangle.
\label{I.04}
\ee
It appears, then, that Eq.~(\ref{I.03}) is the transposition to the level of the observable
${\cal M}$ of the underlying basic relation~(\ref{I.1}). In particular, regarding ${\cal W}$
as the bias which is necessary to apply on $\mu$ in order to construct $\mu^{\prime}$ and ${\cal K}$ as 
the analogous quantity relating $P$ and $P^{\prime}$, Eq.~(\ref{I.04}) tells how these biases
are related one to the other. 

As long as $\mu(\sigma)$ and $\mu^{\prime}(\sigma)$ are arbitrary, the above result does not
have predictive power. 
It becomes the master FT when $\mu^{\prime}(\sigma)$ is taken with a definite relation to $\mu(\sigma)$,
which constrains the form
of $P(M)$ if the right hand side of Eq.~(\ref{I.03}) is accessible without having
to actually compute the expectation, possibly via symmetry arguments.
In the simplest case $\mu^{\prime}(\sigma)$ is taken to be the same as $\mu(\sigma)$. 
Then, from Eq.~(\ref{I.1}) one has
\be
\frac{\mu(\sigma^*)}{\mu(\sigma)} = e^{-{\cal W}(\sigma)},
\label{I.1bis}
\ee
showing that ${\cal W}$ is odd ${\cal W}(\sigma^*) = -{\cal W}(\sigma)$ and
characterizes the asymmetry of the probability measure under inversion. 
Now, the question is to what extent this
asymmetry is preserved or distorted  as the description is moved up
from the microscopic level to the higher one of random variables. The answer from
Eq.~(\ref{I.03}) is given by
\be
\frac{P({\cal M}^* = M)}{P({\cal M} = M)} = e^{-{\cal K}(M)}.
\label{I.3bis}
\ee
Taking ${\cal M}$ to have definite parity, if it is even no information about the symmetry of
$P$ is obtained and Eq.~(\ref{I.3bis}) yields the conditional integral FT
\be
\langle e^{-{\cal W}(\sigma)} | {\cal M}(\sigma) = M \rangle = 1.
\label{I.3tris}
\ee
Conversely, if ${\cal M}$ is odd, one has
\be
\frac{P({\cal M} = -M)}{P({\cal M} = M)} = e^{-{\cal K}(M)},
\label{I.3quater}
\ee
whose usefulness depends on the possibility of assessing the form of ${\cal K}(M)$,
which is called the asymmetry function (AF).
In particular, if ${\cal M} \propto {\cal W}$, there follows
an FT of the Gallavotti-Cohen~\cite{Gallavotti} form with a linear AF
\be
{\cal K}(M) \propto M.
\label{I.3pente}
\ee
If, instead, ${\cal W}$ and ${\cal M}$ are not simply related,
the meaning of ${\cal K}(M)$ in general is not immediately transparent.
For an overview of the variety of the different FT forms arising in the general case see 
Ref.~\cite{Seifert}.

\subsection{FT and symmetry breaking in equilibrium}

The latter remarks are well clarified in the equilibrium context, used
in Ref.~\cite{Gaspard} to investigate the relation between FT and
symmetry breaking.  Let $\Omega$ and $\sigma$ be the system's phase
space and configurations, respectively. For definiteness, let $\sigma
= (s_1,...,s_N)$, with $s_i=\pm 1$, be a spin configuration of a
magnetic system on the lattice in the presence of an external
site dependent field $\mathbb{B}=\{B_i\}$, whose equilibrium state is described
by the probability measure \be \mu(\sigma) = \mu_0(\sigma)e^{\beta
  \sum_i B_i s_i},
\label{equi.1}
\ee
where $\mu_0(\sigma)$ is symmetric under spin inversion $\sigma^* = (-s_1,...,-s_N)$,
while the exponential term breaks explicitly the $\mathbb{Z}_2$ symmetry.
We are interested in the fluctuations of the global magnetization ${\cal M}(\sigma) = \sum_i s_i$.
From Eq.~(\ref{equi.1}) follows
\be
{\cal W}(\sigma) = 2 \beta \sum_i B_i s_i,
\label{equi.01}
\ee
and
\be
{\cal K}(M) = -\ln \langle e^{-2\beta \sum_i B_i s_i}|{\cal M} = M \rangle,
\label{equi.02}
\ee
which takes a simple form only if the external field
is uniform $B_i=B, \forall i$, yielding an FT with the linear AF~\cite{Gaspard}
\be
{\cal K}(M) = 2\beta B M.
\label{equi.3}
\ee
Instead, if $\mathbb{B}$ is not uniform, {\it deviations} from the FT arise.
In order to take a closer look, let the number of spins
to become large and consider the case of the ideal paramagnet, in which
$\mu_0(\sigma)$ in Eq.~(\ref{equi.1}) is the uniform measure $Z^{-1} = [\prod_i 2\cosh (\beta B_i)]^{-1}$.
Then, by a straightforward saddle point
computation one obtains the large deviation principle
\be
P(M) \sim e^{-NI(m)},
\label{G.1}
\ee
where $m=M/N$ is the magnetization per spin and the large deviation function is given by
\be
I(m) = x^*(m) m + \beta \bigl[ f\bigl(\mathbb{B}^*(m)\bigr) - f(\mathbb{B})\bigr].
\label{G.2}
\ee
Here,
\be
f(\mathbb{B}) = \lim_{N \to \infty} - \frac{1}{N\beta} \sum_i \ln [2\cosh (\beta B_i)],
\label{G.3}
\ee
is the Helmotz free energy density, which depends on the field configuration $\mathbb{B}$,
we have defined $\mathbb{B}^*(m)=\{B^*_i(m)\}$ with  
\be
B^*_i(m) = B_i + \beta^{-1} x^*(m),
\label{G.3bis}
\ee
and $x^*(m)$ is obtained by solving with respect to $x$ the
equation of state
\bea
m & = & - \frac{\partial}{\partial x} f\bigl(\{B_i + \beta^{-1} x\}\bigr)  \nonumber \\
& = & \frac{1}{N} \sum_i \tanh (B_i + \beta^{-1} x).
\label{G.4}
\eea
Consequently, $\beta^{-1} x^*(m)$ is the shift to be applied to the external field on each site in order
to produce $m$ as the {\it average} magnetization per spin.
From the definition~(\ref{I.3quater}) follows 
\begin{eqnarray}
& & \frac{1}{N}{\cal K}(M) = I(-m) - I(m) \nonumber \\
& = & - \bigl[x^*(-m) + x^*(m)\bigr]m + \beta \bigl[ f\bigl(\mathbb{B}^*(-m)\bigr) - 
f\bigl(\mathbb{B}^*(m)\bigr)\bigr]. \nonumber \\
\label{G.5}
\end{eqnarray} 
Averaging Eq.~(\ref{G.3bis}) over $i$, we can write
\be
\beta^{-1} x^*(m) = \overline{\mathbb{B}^*(m)} - \overline{\mathbb{B}},
\label{G.01}
\ee
with
\be
\overline{\mathbb{B}^*(m)} = \frac{1}{N} \sum_i B^*_i(m), \,\,\, \overline{\mathbb{B}} = \frac{1}{N} \sum_i B_i,
\label{G.02}
\ee
and Eq.~(\ref{G.5}) can be put in the form
\be
\frac{1}{N}{\cal K}(M) =  2 \beta \overline{\mathbb{B}} m + \beta \bigl[g(-m) - g(m)\bigr],
\label{G.03}
\ee
where we have defined
\be
g(m) = f\bigl(\mathbb{B}^*(m)\bigr) + \overline{\mathbb{B}^*(m)} m.
\label{G.04}
\ee
In the particular case of the uniform external field 
\begin{eqnarray}
& & \overline{\mathbb{B}} = B, \nonumber \\
& & B^*_i(m) = B^*(m) = B + \beta^{-1} x^*(m), \,\,\,\forall i, \nonumber \\
& & \overline{\mathbb{B}^*(m)} = B^*(m), 
\label{G.05}
\end{eqnarray}
and
\be
g(m) = f(B^*) + B^*m
\label{G.06}
\ee
is the Legendre transform of $f(B)$, which is even under $m$ reversal,
since $B^*(m)$ is odd. Hence, in this case Eq.~(\ref{G.03}) reproduces the result~(\ref{equi.3}).
In the nonuniform case $g(m)$ is not the Legendre transform of $f(\mathbb{B})$ and
in general does not have a definite parity, leading to a nonlinear AF function.
We will come back on this point in Section~\ref{twobrownian}.
What the above exercise shows is that the FT, in the sense of a linear AF, holds as
long as the macrovariable ${\cal M}$, whose fluctuations are considered, is conjugate
to the symmetry breaking field. Instead, if ${\cal M}$ is not a conjugate variable, as it is the case
with a site dependent $\mathbb{B}$, the FT in the form~(\ref{equi.3}) does not hold.

\subsection{FT out of equilibrium}

Let us, next, consider the nonequilibrium context. Assuming stochastic
evolution, take for $\Omega$ the space of stochastic trajectories
and for $\sigma$ an individual
trajectory. Then, $\sigma^*$ stands for the time reversed trajectory, while
$\mu(\sigma)$ and $\mu^{\prime}(\sigma)$ are the probability measures associated with
two different evolutions, whose relation is specified from case to case.
Here, as in the equilibrium problem, we shall be concerned with $\mu^{\prime}(\sigma) = \mu(\sigma)$,
which in the dynamical context arises when there are 
no time dependent external parameters and the system evolves
in contact with a single thermal reservoir at the final temperature $T$.
Then, only heat
is exchanged with the environment and one has~\cite{Seifert}
\be
{\cal W}(\sigma) = \ln \frac{P_0(x_0)}{P_0(x_t)} - \beta {\cal Q}(\sigma),
\label{I.001}
\ee
where $P_0$ is the initial probability distribution, $x_0$ and $x_t$ are the
initial and final entries in the trajectory $\sigma$, $\beta = 1/T$ and ${\cal Q}(\sigma)$
is the heat exchanged along the trajectory, which we take as negative if released
to the environment. Using the above form of ${\cal W}$ and the definition~(\ref{I.04}),
the heat AF is given by
\be
{\cal K}(Q) = -\beta Q -\ln \langle e^{\ln P_0(x_t) - \ln P_0(x_0)}|{\cal Q} = Q \rangle,
\label{I.002}
\ee 
whose understanding requires some clue on the role of the boundary
terms. This can be gained from the work of Puglisi {\it et al.}~\cite{Puglisi}. 

In this paper, as anticipated
in the Introduction, we shall be interested in the heat exchanged by a system 
of Brownian oscillators with the environment in an interval of time $[t_w,t]$. Eventually,
this will lead to recast the above equation in the form
\be
{\cal K}(Q) = - \ln \bigl\langle e^{-\sum_{\mathbf{k}} \Delta \beta_{\mathbf{k}} Q_{\mathbf{k}}} |{\cal Q} 
= Q \bigr\rangle,
\label{I.003}
\ee
where $\mathbf{k}$ are single oscillators labels and $Q_{\mathbf{k}}$ the heat exchanged by each one of 
them.
The above expression is clearly analogous to Eq.~(\ref{equi.02}), where the role of
the nonuniform external field $\mathbb{B}$ is played by the set of 
affinities $\{\Delta \beta_{\mathbf{k}} \}$,
which are $\mathbf{k}$-dependent differences of inverse temperatures. The correspondence 
between the two problems helps to understand the deviations or modifications of the FT in terms
of a collection of nonuniform degrees of freedom.

It should be emphasized that the choice of taking $\mu^{\prime} =
\mu$, and therefore $P^{\prime} = P$, is dictated by the particular
physical setting of interest, since we consider the relaxation
following a temperature quench and we want to compare the probability
of exchanging the heat $Q$ with that of exchanging $-Q$, in the {\it
  same quench process}, that is {\it without time reversal}.  We focus
on the asymmetry of the heat distribution in the given process, as it
was done, for instance, in the experimental work of Gomez-Solano et
al.~\cite{gomez11}.

\section{Brownian oscillator}
\label{One}

\noindent 
The equation of motion for the single overdamped Brownian oscillator is of the
Langevin type 
\begin{equation}
\dot{x}=-\omega x + \eta,
\label{process}
\end{equation}
where $\omega$ is the frequency and $\eta$ is the white noise, modeling the
interaction with the thermal bath at the temperature $T$, with expectations
\begin{eqnarray}
\langle \eta(t)\rangle &=&0 \\
\langle \eta(t)\eta(t')\rangle&=&2T\delta(t-t').
\end{eqnarray}
The Boltzmann constant will be taken $k_B=1$ throughout.
Initially the system is in equilibrium at the temperature $T_0$, with the
position probability distribution
\be
P_0(x) = \sqrt{\frac{\beta_0 \omega}{2\pi}} e^{-\beta_0 {\cal H}(x)}, 
\label{HO.1}
\ee
where ${\cal H}(x) = \frac{1}{2}\omega x^2$ is the energy of the oscillator.
Instantaneous cooling (quenches) or heating processes are realized by putting, at the time $t=0$, the system
in contact with the thermal bath at the temperature $T < T_0$ or $T > T_0$, respectively.
In the following, we shall be mainly interested in the case of
the temperature quench.

\subsection{Fluctuations of exchanged heat}

Let us focus on the fluctuations of
the heat exchanged by the oscillator with the thermal bath in the
time interval $(t_w \geq 0,t>t_w)$ after the temperature step. 
Since no work can be carried out on or by the system,
due to $\omega$ constant, the heat exchanged in a single realization of the dynamical evolution
coincides with the energy difference
\be
\mathcal{Q}(t,t_w)={\cal H}\bigl(x(t)\bigr)- {\cal H}\bigl(x(t_w)\bigr),
\label{Phex.1}
\ee
which is positive if heat is absorbed from the bath and negative if
it is released to the bath. Then,
the probability of exchanging the amount $Q$ of heat is given by
\be
P(Q) = \int_{-\infty}^{\infty} dx dx_w \, P(x, t;x_w, t_w) \delta (\mathcal{Q} - Q),
\label{Phex.2}
\ee
where $P(x, t;x_w, t_w)$ is the joint probability of the two events $(x,t)$ and $(x_w,t_w)$,
given by
\begin{eqnarray}\label{HO.9}
P(x, t;x_w, t_w) 
&=&  \frac{1}{\sqrt{2 \pi \Delta(\tau) \nu(t_w)}} \\ \nonumber
&\times & e^{-\left \{ \frac{1}{2\Delta(\tau)}x^2  - \frac{G(\tau)}{\Delta(\tau)} xx_w
+\frac{1}{2}\left [ \frac{G^2(\tau)}{\Delta(\tau)} + \frac{1}{\nu(t_w)} \right ] x_w^2 \right \} },
\end{eqnarray}
where $G(\tau) = e^{-\omega \tau}$
is the response function dependent on the time difference $\tau=t - t_w$,
$\Delta(\tau) = \frac{T}{\omega}[1-G^2(\tau)]$
and $\nu(t_w) = G^2(t_w) \nu_0 + {\Delta(t_w)}$ is the position variance
at the time $t_w$, whose initial value is given by $\nu_0=T_0/\omega$.

The position probability distribution at the generic time $t$, obtained by integrating the
above quantity over $x_w$, maintains the equilibrium form~(\ref{HO.1})
\be
P(x,t) = \sqrt{\frac{\beta_{\rm eff}(t) \omega}{2\pi}} e^{-\beta_{\rm eff}(t){\cal H}(x)},
\label{HO.6}
\ee
where $\beta_{\rm eff}(t)$ is the time dependent inverse effective temperature
defined by the equipartition-like statement~\cite{gomez11} 
\be
\langle {\cal H} \rangle_t = \frac{1}{2}T_{\rm eff}(t),
\label{HO.6bis}
\ee
which yields
\be
T_{\rm eff}(t) =  \omega \nu(t) =  (T_0-T)G^2(t) + T.
\label{HO.10}
\ee
For this quantity, which will play an important role in the following,
we shall use the short hand notation $T_w = T_{\rm eff}(t_w)$ or $T_t = T_{\rm eff}(t)$
and similarly for the inverse temperature $\beta_w = T_w^{-1}$ or $\beta_t = T_t^{-1}$.

Returning to Eq.~(\ref{Phex.2}) and
introducing the integral representation of the $\delta$ function
\be
\delta (\mathcal{Q} - Q) = \int_{-i \infty}^{i \infty} \frac{d \lambda}{2\pi i} \,
e^{-\lambda (Q - \mathcal{Q})},
\label{Phex.3}
\ee
we obtain
\be
P(Q) =  \int_{-i \infty}^{i \infty} \frac{d \lambda}{2\pi i} \, e^{-\lambda Q}
\int_{-\infty}^{\infty} dx dx_w \, \frac{e^{-f(\lambda,x,x_w)}}{2 \pi \sqrt{\Delta(\tau)\nu(t_w)}},
\label{Phex.4}
\ee
where
\be
f(\lambda,x,x_w) = \frac{1}{2 \Delta(\tau)} [x-G(\tau) x_w]^2 + \frac{x_w^2}{2\nu(t_w)}
-\frac{\lambda \omega}{2}(x^2 - x_w^2).
\label{Phex.5}
\ee
Carrying out the $x$ and $x_w$ integrations and rotating the $\lambda$ integration from the
imaginary to the real axis, this can be rewritten as
\be
P(Q) =  \int_{-\infty}^{\infty} \frac{d \lambda}{2\pi \sqrt{a}} \,
\frac{e^{-i \lambda Q}}{\sqrt{(\lambda -i\lambda_-)(\lambda + i\lambda_+)}},
\label{Phex.16bis}
\ee
where
\be
\lambda_\pm = \frac{1}{2 a} \left [\sqrt{b^2 + 4 a} \pm b \right ],\,\,\,\,\lambda_+\lambda_- = 
\frac{1}{a},
\label{Phex.14}
\ee
with
\be
a =  TT_w(1-G^2), \,\,\,\, b =   \Delta T G^2_w(1-G^2),
\label{Phex.11}
\ee
after using the notation
\be
\Delta T = T_0 - T,\,\,\,\, G_w=G(t_w),\,\,\,\,G = G(\tau),
\label{Phex.10bis}
\ee which will be adopted from now on. Let us briefly comment on the
parameters $a$ and $b$ introduced above. The asymmetry
between $\lambda_+$ and $\lambda_-$ is controlled by $b$, while $a$ controls the symmetric
part.  The sign of $b$ depends on that of $\Delta T$, implying
$\lambda_+ > \lambda_-$ in the quench process and $\lambda_+ < \lambda_-$ in the heating process.  
The most asymmetrical situation is
obtained for $a=0$, which is realized either setting $T=0$, or $T_0=0$
and $t_w=0$, yielding 
\be \lambda_+ = \left \{ \begin{array}{ll}
  \infty,\;\; $for$ \;\; b > 0,\\ 1/|b| ,\;\; $for$ \;\; b < 0,
        \end{array}
        \right .
        \label{eig.2}
        \ee
\be
\lambda_-  = \left \{ \begin{array}{ll}
        1/b,\;\; $for$ \;\; b > 0,\\
        \infty,\;\; $for$ \;\; b < 0.
        \end{array}
        \right .
        \label{eig.3}
\ee 
Instead, the symmetrical situation with
\be
\lambda_+ = \lambda_-= \frac{1}{\sqrt{a}},
\label{eig.1}
\ee
is obtained when the system is equilibrated ($b=0$) for $\Delta T=0$ or
for $t_w \rightarrow \infty$.

\begin{figure}[!tb]
\includegraphics[width=0.6\columnwidth,clip=true]{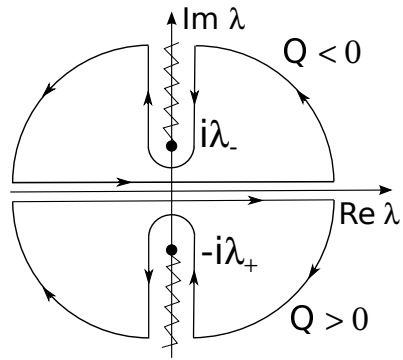}
\caption{Contours of integration, for $Q<0$ and $Q>0$, in the case of a single oscillator.}
\label{fig_int}
\end{figure}

The integral in Eq.~(\ref{Phex.16bis}) is evaluated by closing the contour either on the upper 
half plane or on the lower half plane around the branch cuts
running along the imaginary axis from $\pm i \lambda_{\mp}$ up to $ \pm i \infty$
(see Fig.~\ref{fig_int}), depending on $Q < 0$ or $Q > 0$, and obtaining
\be
P(Q) = \frac{\sqrt{\lambda_+\lambda_-}}{\pi} e^{\frac{1}{2}\Delta \beta Q} 
K_0\left (\frac{\lambda_+ +\lambda_-}{2}|Q| \right ),
\label{Phex.16}
\ee
where 
\be
\Delta \beta = \lambda_- -\lambda_+ = \beta_w - \beta,
\label{aff.1}
\ee
while $K_0$ is the modified Bessel function of the second kind. 

\begin{figure}[!tb]
\includegraphics[width=0.9\columnwidth,clip=true]{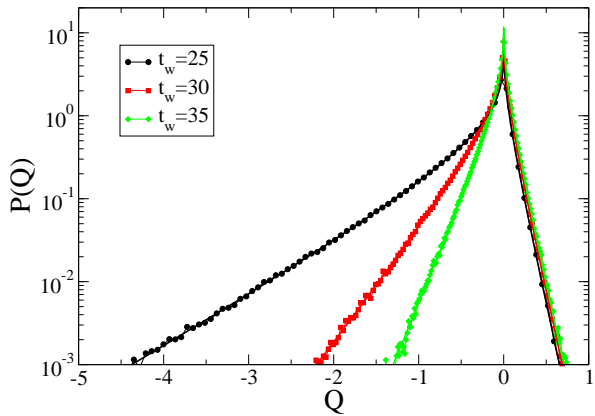}
\caption{Heat probability distribution for the single oscillator, with $\omega=10^{-1}$, $T_0=100$ and $T=10^{-1}$, for $t=200$
and different values of $t_w$.
Analytical form from Eq.~(\ref{Phex.16}) (continuous lines) and
numerical simulations (symbols).}
\label{fig1}
\end{figure}

In the limit $t_w \to \infty$ one has $\lambda_+ = \lambda_- = \beta$, recovering
the equilibrium result
\be
P(Q) = \frac{\beta}{\pi}K_0 (\beta|Q|),
\label{Phex.16ter}
\ee
which was derived in Refs.~\cite{imparato07,chatterjee10} for a Brownian particle
optically trapped in a stationary harmonic potential.  
The complementary case of a Brownian oscillator in the strongly underdamped limit 
has been studied in the recent work of Salazar and Lira~\cite{salazar16}, where a similar
result for the heat distribution is derived. The dependence on $t_w$ of the asymmetry
of the distribution, for a quench to a small but finite temperature $T=0.1$, is illustrated in
Fig.~\ref{fig1}, where the analytical expression~(\ref{Phex.16}) is compared 
with the numerical simulation of the process~(\ref{process}). 

\subsection{Decomposition into cooling and heating contributions}
\label{realization}

The two quantities $\lambda_{\pm}$ turn out to be the basic building blocks in all
what follows.
The physical meaning can be readily understood
by rewriting $P(Q)$ as the convolution of the two distributions arising
in the purely cooling and in the purely heating process. Defining
\be
P_\pm(Q) = \sqrt{\pm i\lambda_\pm} \int_{-\infty}^{\infty} \frac{d \lambda}{2\pi}
\frac{e^{-i \lambda Q}}{\sqrt{\lambda \pm i\lambda_\pm}}
\label{Decomp.1}
\ee
and carrying out the integral, which involves only one of the branch points, one finds
\be
P_-(Q) = \left \{ \begin{array}{ll}
        \sqrt{\frac{\lambda_-}{\pi |Q|}} e^{\lambda_-Q},\;\; $for$ \;\; Q < 0,\\
        0 ,\;\; $for$ \;\; Q > 0,
        \end{array}
        \right .
        \label{Decomp.3}
        \ee
and
\be
P_+(Q)  = \left \{ \begin{array}{ll}
        \sqrt{\frac{\lambda_+}{\pi Q}} e^{-\lambda_+Q},\;\; $for$ \;\; Q > 0,\\
        0 ,\;\; $for$ \;\; Q < 0,
        \end{array}
        \right .
        \label{Decomp.4}
        \ee
from which follows
\be
\lambda_\pm = \pm \frac{1}{2 \langle {\cal Q} \rangle_\pm},
\label{Decomp.5}
\ee
where $\langle {\cal Q} \rangle_\pm$ are the average heats 
exchanged in the processes in which heat can only be absorbed or released.
Then, the probability~(\ref{Phex.16bis}) can be rewritten as 
\be
P(Q) =  \int_{-\infty}^{\min(0,Q)} dQ^{\prime}P_+(Q-Q^{\prime})P_-(Q^{\prime}),
\label{1ho}
\ee
which reproduces the result of Eq.~(\ref{Phex.16}) and
the total average heat exchanged is clearly given by
\bea
\langle {\cal Q} \rangle & = &  \langle {\cal Q} \rangle_+ + \langle {\cal Q} \rangle_- \nonumber \\
& = &
\frac{1}{2} \left ( \frac{1}{\lambda_+} - \frac{1}{\lambda_-} \right ) 
= - \frac{b}{2}.
\label{Decomp.6}
\eea

The physical conditions for $P(Q) = P_-(Q)$ or $P(Q) = P_+(Q)$ are obtained 
putting the oscillator in contact with the thermal reservoir at $T=0$,
or by arranging a purely heating process with $T_0=0, T > 0$ and $t_w=0$.
In the first case, one has
$T_w=T_0 G^2_w, a=0, b=(T_w-T_t)$
from which follows, according to Eqs.~(\ref{eig.2}) and~(\ref{eig.3}),
$\lambda_+=\infty, \lambda_-=(T_w-T_t)^{-1}$, implying
$P_+(Q)= \delta(Q)$ and, therefore, according to Eq.~(\ref{Decomp.3})
\be
P(Q) = P_-(Q) = \left \{ \begin{array}{ll}
        \sqrt{\frac{1}{\pi |Q|(T_w-T_t)}} e^{Q/(T_w-T_t)},\;\; $for$ \;\; Q < 0,\\
        0 ,\;\; $for$ \;\; Q > 0.
        \end{array}
        \right .
        \label{Asy.4}
        \ee
Similarly, in the second case, with $T_0=0$ and $t_w=0$, one has 
$T_w=0, a=0, b=-T_t$, which yield $\lambda_+=1/T_t, \lambda_-=\infty$, 
implying $P_-(Q)= \delta(Q)$ and
\be
P(Q) = P_+(Q) = \left \{ \begin{array}{ll}
        \sqrt{\frac{\beta_t}{\pi Q}} e^{-\beta_tQ},\;\; $for$ \;\; Q > 0,\\
        0 ,\;\; $for$ \;\; Q < 0.
        \end{array}
        \right .
        \label{Asy.8}
        \ee

\subsection{FT and time reversal symmetry breaking}

In this subsection we discuss the symmetry properties of the heat distribution.
From Eq.~(\ref{Phex.16}) follows immediately the FT in the standard form 
\be
\frac{P(-Q)}{P(Q)} = e^{-\Delta \beta Q}.
\label{Phex.21}
\ee
A similar result was derived in Ref.~\cite{salazar16}
with $t_w=0$ and $\Delta \beta = \beta_0 - \beta$, while in Ref.~\cite{gomez11}
there appears $\Delta \beta = \beta_w - \beta_t$, which holds true only in the limit of large
$\tau$ when the system is time decorrelated and $\beta_t \sim \beta$.
Hence, as anticipated in section~\ref{zero}, the AF is linear
\be
{\cal K}(Q) = \Delta \beta Q,
\label{aff.2}
\ee
since the role of the initial probability $P_0$
now is played by $P(x,t_w)$, which after Eqs.~(\ref{Phex.1}) and~(\ref{HO.6})
yields
\be
\ln \frac{P(x_w,t_w)}{P(x,t_w)} = \beta_w {\cal Q}.
\label{aff.3}
\ee

Here, we want to highlight the connection between the above result and the
breaking of the time reversal symmetry, by adapting to the present context
the approach of Gaspard~\cite{Gaspard} sketched in Section~\ref{zero}.
Taking for $\Omega$ the set of ordered pairs $\sigma = (x_w,x)$,
the time reversal symmetry operation is represented by the involution
$\sigma^* = (x,x_w)$ which exchanges the order of $x_w$ and $x$. 
The task is to rewrite the joint probability Eq.~(\ref{HO.9}) in the form of Eq.~(\ref{equi.1}),    
that is as the product of a $\mathbb{Z}_2$ symmetric
factor $\mu_0(\sigma)$ times an exponential where there appears the
heat ${\cal Q}$ in the role of the observable explicitly
breaking the $\mathbb{Z}_2$ symmetry. In order to do this, let us first
cast the joint probability Eq.~(\ref{HO.9}) in the form
\be
\mu(\sigma) = \frac{1}{Z}e^{-{\cal A}(\sigma)},
\label{TRSB.1}
\ee
where the ``action'' ${\cal A}(\sigma)$ can be read out from Eq.~(\ref{HO.9}) as
\be
{\cal A}(\sigma) = \frac{\beta \omega}{2(1-G^2)} \left [x^2 -2Gxx_w + G^2x_w^2 \right ] 
+ \beta_w \frac{1}{2} \omega x_w^2,
\label{TRSB.2}
\ee
and the normalization factor is given by $Z=\sqrt{\frac{2\pi \beta \beta_w \omega^2}{(1-G^2)}}$. 
Decomposing ${\cal A}(\sigma)$ into the sum of an even and an odd part
\be
{\cal A}(\sigma) = {\cal E}(\sigma) + {\cal O}(\sigma),
\label{TRSB.4}
\ee
with
\bea
{\cal E}(\sigma) & = & \frac{1}{2} \left [ {\cal A}(\sigma) + {\cal A}(\sigma^*) \right ] \nonumber \\
& = & \frac{\beta \omega}{4(1-G^2)} \left [(1+G^2)x^2 -4Gxx_w  \right . \nonumber \\
&+&\left . (1+G^2)x_w^2 \right ]
+ \beta_w \frac{1}{4} \omega (x^2 + x_w^2),
\label{TRSB.5}
\eea
and
\be
{\cal O}(\sigma) =  \frac{1}{2} \left [ {\cal A}(\sigma) - {\cal A}(\sigma^*) \right ] 
= -\frac{1}{2} \Delta \beta {\cal Q}(\sigma),
\label{TRSB.6}
\ee
where ${\cal Q}(\sigma)$ is defined by Eq.~(\ref{Phex.1}), 
the form~(\ref{equi.1}) of $\mu(\sigma)$ is obtained
\be
\mu(\sigma) = \mu_0(\sigma) e^{\frac{1}{2} \Delta\beta {\cal Q}(\sigma)},
\label{TRSB.7}
\ee
where
\be
\mu_0(\sigma) = \frac{1}{Z}e^{-{\cal E}(\sigma)},
\label{TRSB.8}
\ee
is $\mathbb{Z}_2$ invariant. Therefore, the invariance under time reversal in
$\mu(\sigma)$ is explicitly broken by the exponential
factor involving ${\cal Q}(\sigma)$ in Eq.~(\ref{TRSB.7}).
Once this is established, the FT in the form of Eq.~(\ref{Phex.21}) 
is recovered straightforwardly. The following
remarks are in order:

\begin{enumerate}

\item  the affinity $\Delta \beta$
driving the heat flow plays the role of the external field
causing the explicit breaking of the $\mathbb{Z}_2$ invariance.

\item  The $\mathbb{Z}_2$
invariant measure $\mu_0(\sigma)$ of Eq.~(\ref{TRSB.8}) is not time translation
invariant, as it could have been naively expected, since the action 
${\cal E}(\sigma)$ in Eq.~(\ref{TRSB.5}) 
depends both on $\tau$, through $G$, and on $t_w$, through $\beta_w$.
Computing $P(Q)$ from Eq.~(\ref{TRSB.7}) one has
\be
P(Q) = \sum_{\sigma} \mu(\sigma) \delta({\cal Q} - Q) = 
e^{\frac{1}{2} \Delta\beta Q} \sum_{\sigma} \mu_0(\sigma) \delta({\cal Q} - Q),
\label{TRSB.9}
\ee
and, comparing with Eq.~(\ref{Phex.16}), one obtains
\be
\sum_{\sigma} \mu_0(\sigma) \delta({\cal Q} - Q) =
\frac{\sqrt{\lambda_+\lambda_-}}{\pi} 
K_0\left (\frac{\lambda_+ +\lambda_-}{2}|Q| \right ).
\label{TRSB.10}
\ee

\item  From Eq.~(\ref{Phex.16bis}) it is straightforward to derive the
moment generating function
\be
\langle e^{\vartheta {\cal Q}} \rangle =
\sqrt {\frac{\lambda_+\lambda_-}{(\vartheta + \lambda_-)(\lambda_+ - \vartheta)}},
\label{CGF.2}
\ee
implying that the cumulant generating function 
$\psi(\vartheta) = \ln \langle e^{\vartheta {\cal Q}} \rangle$
satisfies the same symmetry
\be
\psi(\vartheta) = \psi(-\vartheta + \Delta \beta),
\label{CGF.2bis}
\ee
which was found for a particle coupled to two thermostats at different temperatures~\cite{Farago}.

\end{enumerate}

\section{Two Brownian oscillators}
\label{twobrownian}

Let us, next, see how the overall heat fluctuations are modified when
there are internal degrees of freedom. We consider first the case of
two uncoupled oscillators with frequencies $\omega_1$ and $\omega_2$
and, in the next section, we consider the limit of a large number of
oscillators. 

\begin{figure}[!tb]
\includegraphics[width=0.9\columnwidth,clip=true]{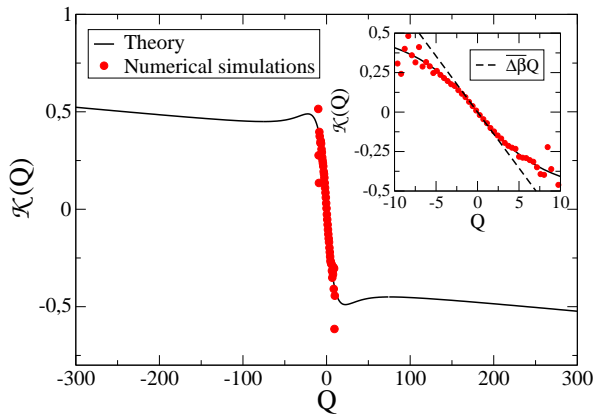}
\caption{Asymmetry function for two Brownian oscillators with
  frequencies $\omega_1=0.5$ and $\omega_2=0.2$, for $T_0=10$, $T=1$,
  $t_w=10$ and $t_w=12$. The line represents the analytical expression
  Eq.~(\ref{moz.4}), while the points are numerical simulations. In
  the inset a zoom of the region at small $Q$ is shown.}
\label{fig_AF}
\end{figure}

As before, initially the system is in equilibrium at the temperature $T_0$ and at the time $t=0$
is put in contact with the reservoir at the temperature $T$. Denoting by
$Q_1$ and $Q_2$ the amounts of heat exchanged by each oscillator,
the probability that the system as a whole exchanges
the heat $Q$ is given by 
\be
P(Q) = \int_{-\infty}^{\infty} dQ_1\, P_1(Q_1)P_2(Q -Q_1),
\label{2osc.1}
\ee
where $P_i(Q_i)$, with $i=1,2$, are the probabilities 
pertaining to each component. 
Inserting the expression~(\ref{Phex.16}) and changing the integration variable from
$Q_1$ to $y = Q_1 - Q/2$, this can be put in the form analogous to Eq.~(\ref{Phex.16})
\be
P(Q) =  e^{\frac{1}{2}\overline{\Delta \beta}Q} R(Q),
\label{moz.1}
\ee
where
\be
\overline{\Delta \beta} = \frac{1}{2}(\Delta \beta_1 + \Delta \beta_2),
\label{moz.01}
\ee
and 
\be
R(Q) = \int_{-\infty}^{\infty} dy \, W_1(y + Q/2) W_2(y - Q/2) 
e^{\frac{1}{2}(\Delta \beta_1 - \Delta \beta_2)y},
\label{moz.2}
\ee
with
\be
W_i(x) = \frac{\sqrt{\lambda_{+,i}\lambda_{-,i}}}{\pi} 
K_0\left (\frac{\lambda_{+,i} +\lambda_{-,i}}{2}|x| \right ).
\label{moz.3}
\ee
Here, $\lambda_{+,i}$ are defined in Eq.~(\ref{Phex.14}), the index $i=1,2$
denoting the oscillator frequencies, $\omega_1$ and $\omega_2$, respectively
and $\Delta \beta_i$ is defined by Eqs.~(\ref{HO.6bis}) and~(\ref{aff.1})
with frequency $\omega_i$.
The resulting AF is given by
\be
{\cal K}(Q) = \overline{\Delta \beta}Q
+ \ln \left (\frac{R(Q)}{R(-Q)} \right ).
\label{moz.4}
\ee 
Hence, if the two oscillators are equal ($\omega_1 = \omega_2$)
the second term disappears, since $R(Q)$ becomes an even function,
and the linear result~(\ref{aff.2}) is recovered.  If they are
unequal, $R(Q)$ is not even under the change of sign of $Q$ and the second term in 
the above equation gives a nonlinear contribution.
The modifications introduced by this second contribution are
illustrated by Fig.~\ref{fig_AF} obtained with parameters
$\omega_1=0.5$ and $\omega_2=0.2$, $T_0=10$, $T=1$ at times
$t=12,t_w=10$.  Around $Q=0$ one observes a linear behavior with slope
$\overline{\Delta \beta}$ (see dashed line in the
inset), while for large values of $Q$ the AF shows a 
nonmonotonic behavior.

In order to expose the connection between the presence of internal structure
and the equilibrium problem mentioned in the Introduction, let us go back
to the microscopic measure.
With two oscillators the phase space enlarges to the set of pairs $\Omega = \{ \sigma_1,\sigma_2 \}$
and the corresponding action, due to the absence of coupling, is additive
\bea
{\cal A}(\sigma_1,\sigma_2) = {\cal A}_1(\sigma_1) + {\cal A}_2(\sigma_2).
\label{FT2.1}
\eea
The two terms in the right hand side are given by Eq.~(\ref{TRSB.2}), keeping into
account that the frequencies $\omega_1$ and $\omega_2$ are different, that is
\bea
{\cal A}_i(\sigma_i) &=& \frac{\beta \omega_i}{2(1-G_i^2)} 
\left [x_i^2 -2G_ix_ix_{w,i} + G_i^2x_{w,i}^2 \right ] 
\nonumber \\
&+& \beta_{w,i} \frac{1}{2} \omega_i x_{w,i}^2.
\label{FT2.2}
\eea
Carrying out the decomposition into even and odd parts, as in Eq.~(\ref{TRSB.4}),
\be
{\cal A}(\sigma_1,\sigma_2) = {\cal E}(\sigma_1,\sigma_2) + {\cal O}(\sigma_1,\sigma_2),
\label{FT2.3}
\ee
the measure can be written as  
\be
\mu(\sigma_1,\sigma_2) = \mu_0(\sigma_1,\sigma_2) e^{-{\cal O}(\sigma_1,\sigma_2)},
\label{FT2.4}
\ee
where 
\be
\mu_0(\sigma_1,\sigma_2) = \frac{1}{Z} e^{-{\cal E}(\sigma_1,\sigma_2)},
\label{FT2.5}
\ee
is even under $* : (\sigma_1,\sigma_2) \mapsto (\sigma^*_1,\sigma^*_2)$, while
\be
{\cal O}(\sigma_1,\sigma_2) = -\frac{1}{2} [ \Delta \beta_1 {\cal Q}_1(\sigma_1) +
\Delta \beta_2  {\cal Q}_2(\sigma_2) ]
\label{FT2.6}
\ee
is odd. Hence, 
${\cal W}(\sigma_1,\sigma_2) = - 2{\cal O}(\sigma_1,\sigma_2)$ and from Eq.~(\ref{I.3quater})
there follows
\be
\frac{P(-Q)}{P(Q)} = e^{-{\cal K}(Q)},
\label{flu.1}
\ee
with
\be
{\cal K}(Q) = -\ln \langle e^{-\sum_i \Delta \beta_i {\cal Q}_i} | {\cal Q} = Q \rangle.
\label{flu.2}
\ee 
The formal
similarity with Eq.~(\ref{equi.02}) is evident, with the affinities $\Delta \beta_i$ playing the role
of the site dependent external field $\{B_i\}$ and the partial heat ${\cal Q}_i$ that of the local
order parameter $s_i$. Accordingly, the above expression simplifies in the particular case 
akin to that of the uniform external field, that is for $\omega_1=\omega_2$ or
for $t_w=0$, yielding
$\Delta \beta_i = \beta_0 - \beta$ for all $i$.

\section{Extended system}
\label{extended}

In this section we consider the case of a large number of oscillators.
The aim is to analyse important qualitatively new features arising in
the behavior of fluctuations when the size of the system becomes
large.  As it is usually the case for large systems, fluctuations of
extensive quantities, such as heat, obey a large deviation
principle~\cite{hugo,ld}. The non trivial feature, unexpected for a
linear system, is that in the large deviation function there appears a
singularity corresponding to a condensation
transition~\cite{Gonnella,Jo,Pechino}. This means, as briefly
anticipated in the Introduction, that there exists a critical
threshold $Q_c$ of the exchanged heat, such that fluctuations above
threshold are due to a {\it single} oscillator.  We give a careful
treatment of the saddle point computation involved in the computation
of the large deviation function and we study the phase diagram of the
transition. Mathematical details are collected in
Appendix~\ref{app:crit}.

We consider an extended system of linear size $L$ and volume $V=L^d$, where $d$ is
the space dimensionality. We suppose that this system is linear and that the normal
modes decomposition with periodic boundary conditions gives rise to a set of harmonic 
components for the low lying modes 
with frequencies obeying the dispersion relation~\cite{note} 
\be
\omega(\mathbf{k}) = k^{\alpha} + \omega_0,
\label{dis.1}
\ee
where $\omega_0 > 0$, $\mathbf{k} = \frac{2\pi}{L} \mathbf{n}, \,\,n_i=0,\pm 1,\pm 2,...$ and $\alpha >0$.
Just to fix ideas, a dispersion relation of this type arises in the Gaussian model
or Van Hove theory of critical phenomena~\cite{Ma}.
The $\mathbf{k}$-space volume per mode is $(2 \pi)^d/V$. Assuming that there exists 
an underlying lattice structure with lattice spacing $a_0$, the first Brillouin zone is bounded by
$\Lambda = 2\pi/a_0$ and the total number of modes is given by $N=V/a_0^d$.
Due to modes independence, the
probability of a generic microscopic state at the time $t_w$ is given by the product
measure
\be
P(\{x_{\mathbf{k}}\},t_w) = \prod_{\mathbf{k}} \sqrt{\frac{\beta_{\mathbf{k},w} \omega(\mathbf{k})}{2 \pi}}
e^{-\beta_{\mathbf{k},w}{\cal H}_{\mathbf{k}}(x_{\mathbf{k}})},
\label{dis.1bis}
\ee
where, according to Eq.~(\ref{HO.10}),
\be
\beta_{\mathbf{k},w} =  [(T_0-T)e^{-2\omega(\mathbf{k})t_w} + T ]^{-1}.
\label{HO.10bis}
\ee
is the inverse effective temperature of the mode $\mathbf{k}$
and ${\cal H}_{\mathbf{k}}(x_{\mathbf{k}}) = (1/2)\omega(\mathbf{k})x^2_{\mathbf{k}}$ is the single mode
energy. Hence, Eq.~(\ref{aff.3}) generalizes to
\be
\frac{P(\{x_{\mathbf{k}}\},t_w)}{P(\{x_{\mathbf{k},w}\},t_w)} = e^{-\sum_{\mathbf{k}} \beta_{\mathbf{k},w} {\cal Q}_{\mathbf{k}}},
\label{dis.1tris}
\ee
where ${\cal Q}_{\mathbf{k}} = {\cal H}_{\mathbf{k}}(x_{\mathbf{k}}) - {\cal H}_{\mathbf{k}}(x_{\mathbf{k},w})$ is the heat
exchanged by the mode $\mathbf{k}$, and the sum is restricted to the first Brillouin zone. 
Inserting into Eq.~(\ref{I.002}) we recover Eq.~(\ref{I.003})
\be
{\cal K}(Q) = - \ln \bigl\langle e^{-\sum_{\mathbf{k}} \Delta \beta_{\mathbf{k}} {\cal Q}_{\mathbf{k}}} |{\cal Q} = Q \bigr\rangle.
\label{dis.001}
\ee
Comparing with Eq.~(\ref{equi.02}), 
we see that the differences $\Delta \beta_{\mathbf{k}}=\beta_{\mathbf{k},w}-\beta$
play the role of the site dependent external field in the equilibrium problem,
as remarked in section~\ref{twobrownian}, and that the AF linear form
is recovered when these differences become independent
of $\mathbf{k}$, as for instance for $t_w=0$, or when the oscillators are
identical, i.e. $\alpha = 0$ (see Appendix~\ref{app:alpha0}).

Let us, next, see what form takes the AF by letting $N$ to become large.
Generalizing Eq.~(\ref{2osc.1}) and keeping $V$ finite,
the probability that the system exchanges the amount of heat $Q$
is given by 
\bea
& & P(Q,V) =  \int_{-\infty}^{\infty} \prod_{\mathbf{k}} dQ_{\mathbf{k}}\, P_{\mathbf{k}}(Q_{\mathbf{k}})\,
\delta(Q - \sum_{\mathbf{k}}Q_{\mathbf{k}}) \nonumber \\
& = & \int_{-i\infty}^{i\infty}\frac{d\lambda}{2\pi i} \, e^{-\lambda Q} 
\prod_{\mathbf{k}} \sqrt{\frac{\lambda_{+}(k)\lambda_{-}(k)}{[\lambda_{+}(k) -\lambda][\lambda_{-}(k) +\lambda]}}, \nonumber \\
\label{dis.3}
\eea
where  $\lambda_{\pm}(k)$ are defined in Eq.~(\ref{Phex.14}),
evaluated for $\omega(\mathbf{k})$ as in Eq.~(\ref{dis.1}). 
Denoting by $\lambda^+_0$ and $\lambda^-_0$ the upper and lower edges of the two branches
of the spectrum, that is 
\be
\lambda^+_0 = \min_{0\leq k\leq \Lambda} \{ \lambda_{+}(k) \}, \,\,\, 
\lambda^-_0 = \min_{0\leq k\leq \Lambda} \{ \lambda_{-}(k) \},
\label{dis.5}
\ee
the contour of integration can be deformed to an arbitrary line $\Gamma$ on the complex plane, provided
it crosses the real axis within the gap $[-\lambda^-_0,\lambda^+_0]$. 
This allows to rewrite the above integral as
\be
P(Q,V) = \int_{\Gamma}\frac{d z}{2\pi i} \, e^{-V[ zq + L(z;V)]},
\label{dis.6}
\ee
where $q=Q/V$ is the density of the exchanged heat and
\be
L(z;V) = \frac{1}{2V}\sum_{0\leq\mathbf{k}\leq\Lambda}
\ln \left[ \frac{[\lambda_{+}(k) - z][z + \lambda_{-}(k)]}{\lambda_{+}(k) \lambda_{-}(k)} \right].
\label{dis.4bis}
\ee
The large $V$ behavior of $P(Q,V)$ can be obtained by 
using the steepest descent method to 
estimate the integral in Eq. \eqref{dis.6}  \cite{nota}.  
When $V\gg1$ 
the wave vector $\mathbf{k}$ can be assumed continuous and 
the sum in Eq. \eqref{dis.4bis}  approximated by an 
integral with an error of $O(1/V)$. 
The neglected $O(1/V)$ terms in the exponent, however, give an
$O(1)$ contribution to $P(Q;V)$,
resulting in an undetermined normalization of $P(Q;V)$,
memory of the discrete nature of $\mathbf{k}$.

The function $L(z;V)$ contains the dangerous boundary terms 
$(1/V)\ln[\lambda_0^{+} - z]$ and
$(1/V)\ln[\lambda_0^{-}+ z]$.
If the steepest descent path $\Gamma$ does not come too close to either edges 
$z = \pm \lambda_0^{\pm}$ of the gap
these terms just give an $O(1/V)$ correction which can be  included into the normalization factor.
Consequently, in this case the function
\be
\label{dis.8}
   L(z) =  \frac{1}{2} \int_0^{\Lambda} d\mu(k) \,
\ln \left[ \frac{[\lambda_{+}(k) - z][z + \lambda_{-}(k)]}{\lambda_{+}(k) \lambda_{-}(k)} \right],
\ee
with 
$d\mu(k) = \Upsilon_d k^{d-1}dk$ and
$\Upsilon_d = [2^{d-1}\pi^{d/2} \Gamma(d/2)]^{-1}$  coming from the
angular integration,
is an uniform approximation to the function $L(z;V)$ as $V\gg 1$.
Here, $\Gamma(x)$ denotes the Euler gamma function.

Denoting by $\phi(z;q) = zq + L(z)$ and by primes derivatives with respect to $z$, 
it is straightforward to verify
that $\mathbb{I}m\, \phi^{\prime}(z;q) \sim \mathbb{I}m\, z$, 
implying that the stationary point $z^*$ of $\phi(z;q)$ is on the real
axis. Thus, setting $z = x + iy$, the coordinate $x^*$ of the saddle point is 
obtained solving the equation $\phi^{\prime}(x^*;q) = 0$,
which explicitly reads
\begin{eqnarray}
q & = &  - L'(x^*) \nonumber \\
& = & \frac{1}{2} \int_0^{\Lambda} d\mu(k) \, \left [ \frac{1}{\lambda_{+}(k) - x^*}
- \frac{1}{\lambda_{-}(k) + x^*} \right ].
\label{Sdlpt.7bis}
\end{eqnarray}
The function $-L'(x)$  is  monotonically increasing in the interval
$x\in [-\lambda_0^-,\lambda_0^+]$, so Eq. \eqref{Sdlpt.7bis} admits solution 
on the condition that $q$ lies between the limiting values:
\be
q_c^{\pm} = -L'(\pm \lambda_0^{\pm}),
\label{Sdlpt.01bis}
\ee
with $q_c^{-} < 0 < q_c^{+}$. 
In this case the steepest descent path in the neighborhood of the saddle point is parallel to the 
imaginary axis
and a straightforward calculation leads to the asymptotic result for $V\gg 1$
\be
\label{eq:PQ_norm}
  P(Q,V) = \frac{1}{\sqrt{-2\pi V L''(x^*)}}\, e^{-V [qx^* + L(x^*)]}.
\ee
If both $q_c^{\pm}$ diverge  this result is valid
for arbitrary finite values of $q$ because $x^* $ is always inside the gap $[-\lambda^-_0,\lambda^+_0]$,
and far from the edges.
If, instead, one or both $q_c^{\pm}$ are finite, the saddle point $x^*$ may reach 
the boundary of the  gap $[-\lambda^-_0,\lambda^+_0]$,
and leave it for $q < q_c^{-}$ or $q > q_c^{+}$.
When this occurs the asymptotic approximation in Eq. \eqref{eq:PQ_norm} is  no more valid.

The boundary values $q_c^{\pm}$ are finite or infinite depending on whether the singularities
of the integrand function in $L'(x)$ are integrable or not.
More specifically, let us denote by $k^{\pm}_0$ the wave vectors at which 
$\lambda_{\pm}(k)$ attain the minimum value,
that is
$\lambda_{\pm}(k^{\pm}_0) = \lambda^{\pm}_0$.
Then, in the neighborhoods of $k^{\pm}_0$, 
\be
\lambda_{{\pm}}(k) - \lambda^{\pm}_0 \sim \begin{cases}
         C\, (k - k^{\pm}_0)^{2n}, & \text{for}\  k^{\pm}_0 > 0;  \\
          C\, k^{\alpha},         & \text{for}\  k^{\pm}_0 = 0,
      \end{cases}
        \label{Sdlpt.10}
\ee
where $C$ is a positive constant,  $n=1,2,\dotsc$ and $\alpha > 0$.
Therefore, $q_c^{\pm}$  diverge if $k^{\pm}_0 > 0$ or if $k^{\pm}_0=0$ and $d \leq \alpha$, while
they are finite if $k^{\pm}_0=0$ and $d > \alpha$. 
Which is the case depends on the parameters of the quench $T_0,T,\tau,t_w$. So, in general,
this manifold of parameters is partitioned into phases
distinguished by $q_c^{\pm}$ being finite or infinite. 

For what follows, and before  discussing how the steepest descent calculation must be modified, it 
is useful to get some insight into the meaning of the two different cases.
Recalling that $\lambda_+(k)$ and
$-\lambda_-(k)$ are the inverse average heat absorbed or released, it is evident
from Eq.~(\ref{Sdlpt.7bis}) that
if $q$ coincides with the average heat density $\langle q \rangle$, then
$x^*(\langle q \rangle) = 0$, implying $-L'(0) = \langle q \rangle$. 
Consequently, if $q$ differs
from $\langle q \rangle$ then $x^*(q) \neq 0$ and 
\be
\lambda^*_{\pm}(k,q) = \lambda_{\pm}(k) \mp x^*(q)
\label{0.02}
\ee
acquire the meaning of the inverse average heat absorbed or released in new conditions, such that
$q$ would be the new average total heat density, exactly as the $\{B_i^*\}$ of section~\ref{zero}
have been recognized to be the shifted external fields necessary to render the fluctuation
$m$ equal to the average magnetization per spin.
In other words, a deviation of $q$ 
from the average $\langle q \rangle$
brings in a shift by $x^*(q)$ of the inverse heats exchanged by the single modes. 
Accordingly, when $x^*(q)$ approaches the edges of the gap the above defined $\lambda_0^{\pm *}$ vanish,
which means that the corresponding edge modes give an {\it infinite} contribution to the exchanged heat.
This may happen either because the fluctuation $q$ itself is infinite, or because the contributions
of all the modes, other than the edge ones, can sum up at most to the finite amounts $q_c^{\pm}$. In the
latter case, if $|q|$ goes above the thresholds $|q_c^{\pm}|$, in order to make up
for the finite difference $|q| - |q_c^{\pm}|$ a spike contribution must come from the edge modes
at $k_0^{\pm}$. This phenomenon is the condensation of fluctuations
in the edge mode, since as anticipated in the Introduction and at the beginning of
this section, the entire amount of
the fluctuation above the critical threshold comes from a single degree of freedom. 
The mechanism of the transition can be recognized to be the same as that of 
Bose-Eintein condensation~\cite{Huang}.
More in general, condensation of fluctuations
has been found in widely different contexts such as information theory~\cite{Merhav},
finance~\cite{Filiasi} and statistical mechanics encompassing
both equilibirum and out of equilibrium situations~\cite{JoEPL}.

When one or both $q_c^{\pm}$ are finite, the steepest descent calculation must be 
modified.
This case will be dealt with in the next subsection. Returning to the case in which
Eq.~(\ref{eq:PQ_norm}) holds, neglecting subdominant terms and taking the ratio $P(-Q)/P(Q)$, 
one obtains the following explicit form of the AF
\be
\frac{1}{V}{\cal K}(Q) = -[x^*(-q) + x^*(q)]q + [L(x^*(-q)) - L(x^*(q))],
\label{condns.7}
\ee
which is clearly reminiscent of Eq.~(\ref{G.5}) in sect.~\ref{zero}. 
In order to expose the analogy, let us
subtract equations~(\ref{0.02}) one from the other and take the average over $k$,
obtaining
\be
x^*(q) = \frac{1}{2} \left (\overline{\lambda_+} - \overline{\lambda_-} \right )
+ \frac{1}{2} \left [\overline{\lambda_-^*(q)} - \overline{\lambda_+^*(q)} \right ],
\label{0.03}
\ee
where 
\be
\overline{\lambda_{\pm}} = \frac{1}{{\cal N}}\int_0^{\Lambda} d\mu(k) \, \lambda_{\pm}(k), \,\,\,
{\cal N} = \int_0^{\Lambda} d\mu(k) \, 1,
\label{0.04}
\ee
and similar expressions for $\overline{\lambda_{\pm}^*(q)}$. Inserting into Eq.~(\ref{condns.7}),
we obtain
\be
\frac{1}{V}{\cal K}(Q) = \left (\overline{\lambda_-} - \overline{\lambda_+} \right )q
+ \left [\Psi(-q) - \Psi(q) \right ],
\label{0.05}
\ee 
where
\be
\Psi(q) = L(x^*(q)) + 
\frac{1}{2} \left [ \overline{\lambda_-^*(q)} - \overline{\lambda_+^*(q)} \right ]q.
\label{0.06}
\ee 
Hence, comparing with Eqs.~(\ref{G.03}) and~(\ref{moz.4}), we can recognize the same structure: 
the prefactor $\left (\overline{\lambda_-} - \overline{\lambda_+} \right )$ 
of the linear contribution (the same appearing in the single oscillator case)
plays the role of the external average field $\mathbb{B}$, while the nonlinear
term $\left [\Psi(-q) - \Psi(q) \right ]$, arises from ``free energy'' contributions.
As remarked above, the presence of this latter term is due to the $k$-dependence of $\Delta \beta_k$,
which plays the same role as the $i$-dependence of $B_i$.

\subsection{Condensation transition}

On physical grounds one can argue that condensation at finite
$q_c^{-}$ can occur only in cooling experiments, where the final
temperature is lower than the initial one, while condensation at
finite $q_c^{+}$ can occur only in heating experiment.  We do not have
a proof of this, but numerical analysis of the conditions
for condensation confirm this conjecture.

Let us thus concentrate on the case with $q_c^{+} = \infty$ and
$q_c^{-}$ finite, which occurs for $k_0^- = 0$ and $d > \alpha$.  The
analysis of the opposite case in which $q_c^{+}$ is finite and
$q_c^{-} = -\infty$, (or both finite, if such a case exists) is
straightforward.
\begin{figure}[!tb]
\includegraphics[width=0.9\columnwidth,clip=true]{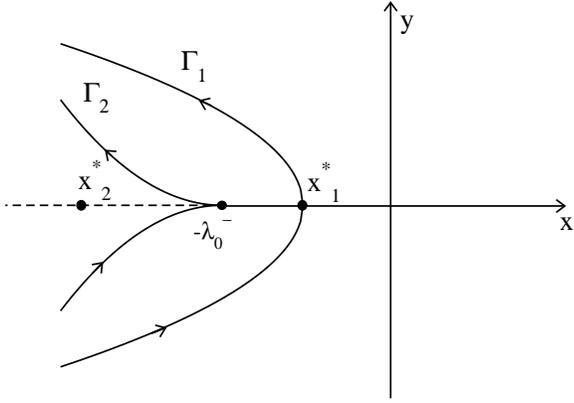}
\caption{Schematic representation of the steepest descent path of
  integration. If $x^*=x_1^*\ge -\lambda_0^-$, the integration contour
  goes through $x_1^*$, where $\phi'=0$, ($\Gamma_1$ curve).
  If $x^*=x_2^*< -\lambda_0^-$,
  the integration contour develops a cusp and sticks
  to $-\lambda_0^-$, ($\Gamma_2$ curve).}
\label{cusp}
\end{figure}
When $q_c^-$ is finite,  Eq.~(\ref{Sdlpt.7bis}) 
does not admit a solution with  $x^*\in[-\lambda_0^{-}, \lambda_0^{+}]$ whenever $q < q_c^-$.
The problem is well known from the theory of Bose-Einstein
condensation of an ideal gas of bosons~\cite{Huang} or from
the ``sticking'' of the saddle point to a singularity
in the solution of the spherical model of ferromagnetism~\cite{Berlin}. 
It arises because 
when $q < q_c^{-} = -L'(\lambda_0^{-})$ the steepest descent path cannot pass through the 
stationary point $x^*< -\lambda_0^{-}$ of $\phi(z;q) = zq + L(z)$ 
and must traverse the real axis at the gap edge $z=-\lambda_0^{-}$. 
Since $\phi'(-\lambda_0^-;q) = q - q_c^{-}$
is negative for $q < q_c^{-}$,  the steepest descent path at $z=-\lambda_0^{-}$ bends 
toward the negative real axis forming  
the cusp characteristic of saddle point sticking at the gap edge (see Fig. \ref{cusp}).
The large $V$ behavior of $P(Q,V)$ is dominated by the 
neighborhood of the gap edge because $\phi(-\lambda_0^{-};q) < \phi(x^*;q)$
leading to  the asymptotic behavior for $V\gg 1$:
\be
\label{eq:PQ_cond1}
  P(Q,V) 
  = \frac{1}{\sqrt{-\pi V \bigl(q - q_c^{-}\bigr)}}\,
         e^{-V[ -\lambda_0^{-} q + L(-\lambda_0^{-})]},
\ee
valid for $q < q_c^{-}$.

As a matter of fact, this expression and Eq. \eqref{eq:PQ_norm} valid
for $q > q_c^{-}$, hold for $x^*$ not to close to the gap edge
$-\lambda_0^{-}$.  That is, respectively, in the condensed phase and
normal phase for $q$ not too close to $q_c^{-}$.  The analysis of the
asymptotic behavior of $P(Q,V)$ as $V\gg 1$ for all values of $q$,
including close to $q_c^{-}$, requires some care.  Details can be
found in Appendix \ref{app:crit}.

\begin{figure}[!tb]
\includegraphics[width=0.9\columnwidth,clip=true]{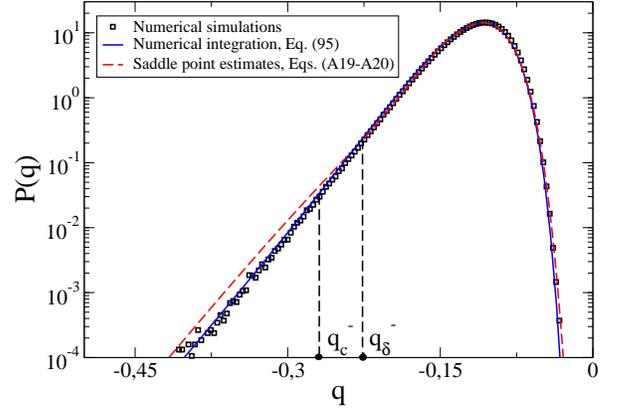}
\caption{Comparison of the saddle point computation of $P(q)$ with the
  numerical computation of the same quantity for $\omega_0=0.1$, $T_0=100$, $T=0.1$,
  $t_w=3$, $\tau=7$, for a system with $L=41$ in $d=2$ ($N=1681$
  oscillators), and with $\alpha=1$. (The quantity $q_\delta^-$ is defined in Eq.~(\ref{eq:q_d})).}
\label{fig20}
\end{figure}

Adding and subtracting $\lambda_0^{-}q_c^-$ in the exponent of Eq. \eqref{eq:PQ_cond1}, 
$P(Q;V)$ in the condensed phase far from the critical point can be written as
\be
  P(Q,V) =   \frac{e^{\lambda_0^- (Q-Q_c^-)}}{\sqrt{\pi \bigl|Q - Q_c^{-}\bigr|}}  \,
e^{V[\lambda_0^- q_c^-  - L(-\lambda_0^-)]},
\label{Img.21}
\ee 
where $Q_c^- = Vq_c^-$.
Comparing the first term with the expression~(\ref{Decomp.3})
for the distribution in the purely cooling process we have
\be
   P(Q,V) =  P_{k=0,-}(Q-Q^-_c)\, P(Q^-_c,V),
\label{Img.22}
\ee
where, up to normalization factors, 
\be
  P(Q_c^-,V)  \sim    e^{V [\lambda_0^{-} q_c^{-} - L(-\lambda_0^{-})]}.
\ee
This means that negative fluctuations below the critical lower threshold $Q^-_c$
condense into the cooling contribution of the $\mathbf{k} =0$ mode for the exceeding
part $(Q - Q^-_c)$, while the contribution of all the other modes is locked onto
$Q^-_c$.
The comparison of the saddle point estimates of $P(Q,V)$, in the normal and in the condensed
phase, with the ``exact'' numerical computation is illustrated in Fig.~\ref{fig20}.

\subsection{Phase diagram}

\begin{figure}[!tb]
\includegraphics[width=0.9\columnwidth,clip=true]{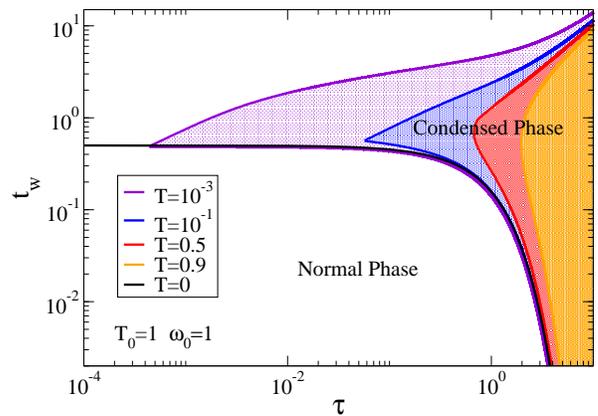}
\caption{$q_c^-$ phase diagram for various final temperatures $T$ and
for $T_0=1$. The normal phase corresponds to $q_c^- = \infty$, while
the condensed phase to $q_c^-$ finite. }
\label{Phasediagr}
\end{figure}

In order to establish the ``phase'' structure, it is necessary to analyse
the behavior of $k^{\pm}_0 > 0$. Let us first recall the results 
of Ref.~\cite{Pechino}, where the problem was analysed for the quench
to $T=0$. This is the simplest case because,
as specified in section~\ref{realization}, heat can only be released
and we have to deal only with the $\lambda_-(k)$ branch of the spectrum.
Differentiating with respect to $k$ the expression for $\lambda_-(k)$, given by the 
first line of Eq.~(\ref{eig.3}), we get
\be
\frac{\partial \lambda_-}{\partial k} = C(k,\tau,t_w)A(k,\tau,t_w),
\label{Part.01}
\ee
where $C(k,\tau,t_w) = b^{-2} \Delta T e^{-2\omega(\vec k)(t_w + \tau)}\alpha k^{\alpha -1}$
is a positive quantity, while 
\be
A(k,\tau,t_w) = t_w e^{2\omega(\vec k)\tau}[1-e^{2(E-k^{\alpha})\tau}],
\label{Part.02}
\ee
has the sign of $(k^{\alpha}-E)$, with 
\be
E(\tau,t_w) = \frac{1}{2\tau} \ln \left (1 + \frac{\tau}{t_w} \right ) - \omega_0,
\label{Part.03}
\ee
and vanishes at $k^{\alpha} = E$.
Therefore, imposing the condition $E(\tau,t_w) = 0$, in the $(\tau,t_w)$ plane
there remains defined the critical line, given by
\begin{equation}
t_w = \frac{\tau}{e^{2\omega_0 \tau} - 1},
\label{Part.6}
\end{equation}
such that above it $k_0^- =0$, while below $k_0^- > 0$. Thus, $k_0^-$ or, equivalently,
$q_c^-$ act as order parameters and below the critical line the system is in the 
normal phase, corresponding to $q_c^-=-\infty$, while above it is in the 
condensed phase, corresponding to $q_c^-$ finite,
as illustrated in Fig.~\ref{Phasediagr} (black line).

If $T > 0$, we must take into account both branches $\lambda_{\pm}(k)$ 
and keep track of the two order parameters $q_c^{\pm}$. The
analytical search of the critical lines turns out to be quite complicated. 
So, we have looked for the absolute minima of $\lambda_{\pm}(k)$ numerically.
The resulting phase diagram for $q_c^-$ is 
depicted in Fig.~\ref{Phasediagr}  for a few values of $T$, ranging from very low to
almost equal to $T_0$. When the system is equilibrated ($T=T_0$), there is no
condensed phase. Fig.~\ref{Phasediagr} shows that upon lowering $T$ there appears a condensed
phase which, starting from the far right, pronges toward the left eventually filling the entire
region above the $T=0$ critical line. The prominent qualitative difference between the $T=0$
and the $T>0$ phase diagrams for $q_c^-$ is that in the latter case the transition driven by an increasing
$t_w$, for fixed $\tau$, manifests reentrant behavior. The same computation for
$q_c^+$ does not show the existence of a condensed phase, namely $q_c^+ = \infty$ all
over the explored $(\tau,t_w)$ plane.

The reentrant behavior of the phase diagram can be interpreted as follows. Let us consider a quench at finite
temperature, and let us fix $\tau$. Then,
for small enough $t_w$, all oscillators are out of equilibrium and can exchange an arbitrary amount
of heat with the bath. Therefore no condensation can take place in this region. Upon increasing $t_w$,
all oscillators do equilibrate but the slowest one, corresponding to the mode $k=0$, 
that can account for large exchange of heat (above the threshold), leading to the condensation of fluctuations. Eventually,
for large $t_w$, all oscillators are equilibrated and again the normal phase is recovered.

\section{Conclusions}
\label{conclusions}

In this paper we have studied the fluctuations of the heat exchanged with the environment
by a system of oscillators relaxing after a temperature quench. Focusing on the
FT, we have investigated the relation between the deviations from linearity of the AF
and the internal structure of the system, building on the analogy with the behavior
of fluctuations in equilibrium when a symmetry is broken by nonuniform external
perturbations.

We have first analysed in great detail the case of a single
oscillator, reducing it to the convolution of the two elementary
processes in which heat can be only released or only absorbed. The
inverse average heats released $(\lambda_-)$ and absorbed
$(\lambda_+)$ in these processes turn out to be the basic objects
underlying the behavior of the quantities of interest.  In the
one-oscillator case the AF is linear with slope given by the
difference $\Delta \beta = \lambda_- - \lambda_+$ and, in the
framework of the above mentioned analogy with the equilibrium problem,
this quantity acts like an external field explicitly breaking the time
reversal symmetry. By adding a second oscillator it starts to surface
the role of the dishomogeneity of the external perturbation in
determining deviations from linearity in the AF.  It should be pointed
out that in the dynamical problem the lack of homogeneity is due to
the existence of different relaxation rates, related to the presence
of degrees of freedom evolving on different time scales.  This
produces a temporal dishomogeneity, which induces a differentiation of
the affinities $\Delta \beta_1$ and $\Delta \beta_2$, analogous to the
spatial heterogeneity generated by the external field $\{B_i\}$.

This picture emerges most clearly in the case of a large number of
oscillators. In this case, the system, although diagonal, presents
nontrivial features, the most notable of which is the possible
condensation of fluctuations~\cite{Gonnella,Jo,Pechino}. Here, we have
presented a detailed analytical study of the large deviation function
of the heat distribution, with an accurate treatment of the crossover
from the normal to the condensed phase, pointing out some interesting
technical points involved in the application of the steepest descent
method in the presence of a transition. In addition, we have mapped
out the phase diagram of the condensation transition, discovering an
unexpected reentrant behavior, in the $(t_w,\tau)$ plane.

Our analysis allowed us to establish a close correspondence with the
structure of the AF for the paramagnet in equilibrium under the action
of a nonuniform external field.  In the equilibrium case it seems
reasonable to make the statement that the linearity of the AF depends
on whether the observable of interest ${\cal M}(\sigma)$ is conjugate
to the external perturbation, the implication of which being that
deviations from the FT are due to the lack of conjugation. Clearly,
the interesting and challenging issue is to understand whether this
way of looking at the AF and its deviations from linearity can be extended to the
nonequilibrium case. The study we have presented in this paper is a
first step in this direction, paving the way to futures investigations
within the context of more complex interacting systems, showing slow
relaxation and aging phenomena.

\appendix

\section{}
\label{app:crit}

In this Appendix we derive the uniform asymptotic expansions 
of  $P(Q,V)$ as $V\gg 1$ valid for all values of  $q=Q/V$, including at
the condensation transition. 
As in the main text we shall assume $q_c^{+} = \infty$ and
$q_c^{-}$ finite so that the steepest descent path hits the gap lower edge 
at the finite value $q = q_c^{-}$.
Close to the edge $z=-\lambda_0^{-}$ the dangerous boundary term 
$(1/V)\ln[\lambda_0^{-}+ z]$ is no more negligible,
thus the uniform approximation $L(z)$,
Eq. \eqref{dis.8}, to $L(z;V)$ as $V\gg 1$ breaks up and shuld be replaced by
\be
\label{eq:G_1}
  L(z;V) = 
    \frac{1}{2V} \ln \bigl[\lambda_0^{-} + z\bigr] + L(z) + O(1/V).
\ee
However, the function $L(z)$ is nonanalytic  at $z=-\lambda_0^{-}$ while
$L(z;V) - \frac{1}{2V} \ln \bigl[\lambda_0^{-} + z\bigr] $ is regular at the gap edge 
$-\lambda_0^{-}$, see Eq. \eqref{dis.4bis}.
Therefore, to construct an uniform asymptotic  approximation to $P(Q;V)$ as  $V\gg 1$ valid for all 
values of $q$ the function $L(z)$ must be regularised. 
Hence, in Eq. \eqref{eq:G_1} the function $L(z)$ is replaced by
\be
\label{eq:phi_delta}
\begin{split}
  L_\delta(z) &=
     \frac{1}{2}\int_{\delta}^{\Lambda} d\mu(k) \ln\left[\frac{\lambda_-(k) + z}{\lambda_-(k)}\right]
          \\ 
     &\phantom{====}
               +\frac{1}{2}\int_{0}^{\Lambda} d\mu(k) \ln\left[\frac{\lambda_{+}(k) - z}{\lambda_{+}(k)}\right],
\end{split}     
\ee
where $\delta = O(1/V^{1/d})$ is an infrared cut-off \cite{Note2}. 
Without loss of generality we can take $\delta = V^{-1/d}$  because any proportionality constant 
can be absorbed into the $O(1/V)$ 
corrections in Eq. \eqref{eq:G_1}.
Notice that the boundary value of $q$ separating the two phases becomes:
\be
\label{eq:q_d}
  q_\delta^{-} = - L_\delta'(-\lambda_0^{-}).
\ee
Using $\lambda_{+}(k) - \lambda_0^{-} \sim C k^\alpha$  as $k \ll 1$, 
the finite volume correction to the critical point reads in the large $V$ limit:
\be
  q_\delta^{-} = q_c^{-} + \frac{\Upsilon_d}{2C(d-\alpha)} V^{\alpha/d -1} + o(V^{\alpha/d -1}),
\ee
where $\Upsilon_d = [2^{d-1}\pi^{d/2} \Gamma(d/2)]^{-1}$ and $\Gamma(x)$ is the Euler's gamma 
function.

Replacing $L(z;V)$ into  Eq. \eqref{dis.6}, 
neglecting the contribution from the $O(1/V)$ terms
and taking  $z=-\lambda_0^{-} + t$ 
to move the end point of the cut $(-\infty;-\lambda_0^{-}]$ on the negative real axis 
to $t=0$, leads to:
\be
\label{eq:app_PQ}
    P(Q,V) = \int_{\Gamma} \frac{dt}{2\pi i}\, t^{-1/2}\, 
         e^{-V \phi(-\lambda_0^{-}+ t;q)},
\ee
where $\phi(z;q) = zq + L_\delta(z)$. 
As discussed in the main text, the stationary point $t^*$ of $\phi(-\lambda_0^{-}+t;q)$, solution of 
$\phi'(-\lambda_0^{-}+t^*;q) = q + L_\delta'(-\lambda_0^{-}+t^*) = 0$ is on the real axis.
The function $\phi(-\lambda_0^{-}+ t;q)$ is analytic at $t=0$ and
\be
\label{eq:app_phip}
  \phi_\delta'(-\lambda_0^{-}; q) = q + L_\delta'(-\lambda_0^{-}) = q - q_\delta^{-}
\ee
changes sign at  $q=q_\delta^{-}$, and 
is  negative if $ q < q_\delta^{-}$ and positive  if $q > q_\delta^{-}$.

Far from the critical point, i.e., 
$|q-q_\delta^{-}| = O(1)$ as $V\gg 1$, 
 the steepest descent calculation is straightforward.
 If $q > q_\delta^{-} $ the saddle point lies on the positive $t$-axis and the steepest descent 
 path can pass through it. Near $t^*$ the steepest descent path is parallel to the imaginary 
 axis and  
 \be
\label{eq:app_PQ_norm_out}
  P_{\rm out}(Q,V) = \frac{e^{-V \phi(-\lambda_0^{-}+ t^*;q)}}
                                 {\sqrt{-2\pi V t^* L_\delta''(-\lambda_0^{-} + t^*)}},
\ee
 as $V\gg 1$, cfr.  Eq. \eqref{eq:PQ_norm}.
 In the opposite case $q  < q_\delta^{-}$ the stationary point $t^*$ is negative and 
 the steepest descent path cannot pass through it.
Thus it must traverse the real axis at the gap edge $t=0$. 
Since $\phi_\delta'(-\lambda_0^-;q)$
is negative for $q < q_\delta^{-}$,  the steepest descent path bends 
toward the negative real axis at $t=0$  forming  
the cusp characteristic of saddle point sticking at the gap edge (see Fig. \ref{cusp}).
The asymptotic behavior for $V\gg 1$ is dominated by the gap edge because 
$\phi(-\lambda_0^{-};q) < \phi(-\lambda_0^{-}+t^*;q)$. 
In the neighborhood of $t=0$ the steepest descent path is parallel to the negative real axis,
hence using \eqref{eq:app_phip}, 
 \be
 \label{eq:app_PQ_cond_out}
  P_{\rm out}(Q,V) 
      = \frac{e^{-V\phi(-\lambda_0^{-}; q) }}{\sqrt{-\pi V \bigl(q - q_\delta^{-}\bigr)}},
\ee
as $V\gg 1$,  cfr. Eq. \eqref{eq:PQ_cond1}.

When $|q - q_\delta^{-}| \ll 1$ as $V\gg1$, $q$ is very close to $q_\delta^{-}$ and the stationary 
point is at
$|t^*| \ll 1$. In this region we can expand $\phi(-\lambda_0^{-} + t; q)$ in powers of $t$:
\be
 \label{eq:app_phi}
  \phi(-\lambda_0^{-} + t; q)= \phi_0 + \phi_0'\, t + \frac{1}{2}\phi_0''\, t^2 + O(t^3), 
  \quad |t|\ll 1,
\ee
where $\phi_0^{(n)} = \phi^{(n)}(-\lambda_0^{-}; q)$.
The first three terms of the expansion must be retained; subsequent terms just give corrections to
the leading asymptotic expansion and can be neglected. 

Using the expansion \eqref{eq:app_phi} the stationary point is at:
\be
  \phi'(-\lambda_0^{-}+t^*; q) = 0 \ \Rightarrow\ 
      t^* = - \frac{\phi_0'}{\phi_0''}
           = - \frac{q - q_\delta^{-}}{L_\delta(-\lambda_0^{-})}.
\ee

Consider first the case $\phi_0' > 0$, that is $q > q_\delta^{-}$. In this case $t^*>0$, because 
$L_\delta''(-\lambda_0^-) < 0 $, and the steepest descent path can go through the saddle point 
at $t^*$.  Thus, expanding Eq. \eqref{eq:app_phi} around $t^*$ and inserting the results into  
Eq. \eqref{eq:app_PQ} leads to:
\be
\label{eq:app_PQ_P}
  P_{\rm in}(Q,V) =  \norm_{+}(q)\, \exp\left[-V \left(\phi_0 - \frac{\phi_0'^2}{2 \phi_0''} \right)\right]
\ee
where
\be
\label{eq:app_NP}
  \norm_{+}(q) = \int_{\Gamma} \frac{dt}{2\pi i}\, t^{-1/2}\, 
         e^{\frac{V}{2} |\phi_0''|  (t-t^*)^2}.
\ee
The steepest descent path  is  the vertical line $\mathbb{R}e\, (t-t^*) = 0$ passing at $t^*$.
Inside the critical region 
$|\phi_0''|(t-t^*)^2 = O(1/V)$ as $V\gg1$, so  
the steepest descent path is given by $t-t^* = i y /\sqrt{V|\phi_0''|}$. Introducing the variable
\be
\label{eq:app_xi}
  \xi = \sqrt{V|\phi_0''|}\, t^* = \sqrt{\frac{V}{|\phi_0''|}}\, \phi_0',
\ee
Eq. \eqref{eq:app_NP} becomes:
\be
   \norm_{+}(\xi) =  \frac{1}{[V|\phi_0''|]^{1/4}} 
                       \int_{-\infty}^{+\infty}  \frac{dy}{2\pi}\, (\xi + i y)^{-1/2} e^{-y^2/2}. \\
\ee
To evaluate the integral we shift the integration axis vertically by $i\xi$. The singularity $(\xi + i y)^{-1/2}$ at $y=i\xi$  gives no contribution,
so we have
\begin{align}
\label{eq:app_NP_D}
   \norm_{+}(\xi) &= \frac{1}{2\pi} \frac{e^{\xi^2/2}}{[V|\phi_0''|]^{1/4}} 
                       \int_{-\infty}^{+\infty}  dy\, (i y)^{-1/2} e^{-y^2/2 - i\xi y} 
\nonumber                        \\
                       &= \frac{1}{\sqrt{2\pi}} \frac{e^{\xi^2/4}\, D_{-1/2}(\xi)}{[V|\phi_0''|]^{1/4}},
\end{align}
where $D_\nu(\xi)$ are the parabolic cylinder functions~\cite{grad}.
Thus, collecting all terms,
\be
\label{eq:app_PQ_in_norm}
  P_{\rm in}(Q,V) = \frac{1}{\sqrt{2\pi}} \frac{e^{-\xi^2/4} D_{-1/2}(\xi)}
                                   {[V|L_\delta''(-\lambda_0^{-})|]^{1/4}}\, 
    e^{-V\phi(-\lambda_0^{-};q)}.
\ee
In the limit $\xi\gg 1$, 
using the asymptotic expansion $D_{-1/2}(\xi) \sim e^{-\xi^2/4}\, /\sqrt{\xi}$ valid for $\xi \gg 1$, 
Eq. \eqref{eq:app_NP_D}   becomes:
\be
\label{eq:app_PQ_match_norm}
  P_{\rm match}(Q,V) =  \frac{e^{-V \phi_0 +  \frac{\xi^2}{2} }}
             {\sqrt{2\pi V (q - q_\delta^{-})}},
\ee
which match asymptotically with the $q - q_\delta^{-} \ll 1$ limit of 
Eq. \eqref{eq:app_PQ_norm_out}.
Then,
\be
 \label{eq:app_PQ_unif}
  P(Q;V) = \frac{P_{\rm in}(Q,V)\, P_{\rm out}(Q,V)}{P_{\rm match}(Q,V)}
\ee
gives an uniform asymptotic approximation to Eq.~\eqref{eq:app_PQ} as $V\gg1$ valid for 
$q \geq q_\delta^{-}$. 
Using Eqs. \eqref{eq:app_PQ_norm_out}, \eqref{eq:app_PQ_in_norm} and 
\eqref{eq:app_PQ_match_norm},
we have
\be
\label{eq:app_PQ_norm_unif}
  P(Q;V) = {\cal N}_{+}(\xi) 
                   \sqrt{\frac{q - q_\delta^{-}}{t^* |L_\delta''(-\lambda_0^{-}+t^*)|}}\,
                   e^{-V \phi(-\lambda_0^{-} + t^*;q)}.
\ee

\begin{figure}[!tb]
\includegraphics[width=0.9\columnwidth,clip=true]{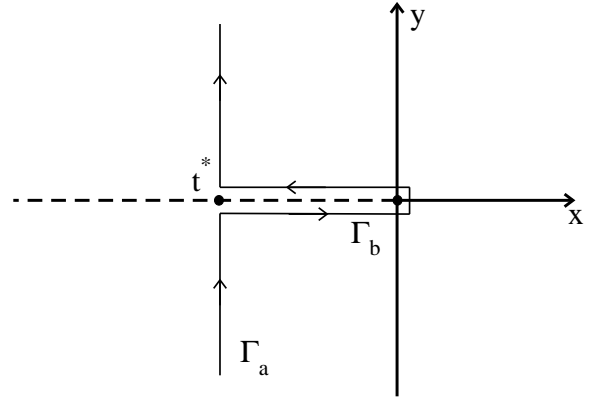}
\caption{Steepest descent path in Eq.~(\ref{A}).}
\label{path_app}
\end{figure}

If, instead, $q < q_\delta^{-}$ then $\phi' <0$ and $t^* <0$. The stationary point lies now on the 
cut $\mathbb{R}e\, t < 0$ and the steepest descent path cannot pass through it. 
Substituting the expansion \eqref{eq:app_phi} into Eq. \eqref{eq:app_PQ} leads to:
\be
 \label{eq:app_PQ_M}
  P(Q,V) =  \norm_{-}(q)\, e^{-V \phi_\delta(-\lambda_0^{-};q)},
\ee
where 
\be
  \norm_{-}(q) = \int_{\Gamma} \frac{dt}{2\pi i}\, t^{-1/2}\, 
         e^{-V \bigl[  \phi_0'\, t + \frac{1}{2}|\phi_0''|  t^2\bigr]}.
\label{A}
\ee
Taking $t = x + iy$ the equation of the steepest descent path  reads
$y (\phi_0' + \phi_0'' x) =0$. 
The steepest descent path is therefore composed by: 
$\Gamma_a$) the two vertical paths $x = -\phi_0'/\phi_0'' = t^*$ , i.e., the steepest descents paths 
from either side of the saddle point;
$\Gamma_b$)  the two paths $y=0$  on either side of the cut joining the saddle 
point $t^*$ with the point $t=0$, where the path can cross the real axis (see Fig.~\ref{path_app}).

The point $t=0$ lies on the steepest ascent path issuing from the saddle point at $t^*$ and hence
it will dominate the integral as $V\gg 1$ whenever $t^* = O(1)$ as $V \gg 1$.  In this case only
the paths $\Gamma_b$ contribute and a straightforward calculation leads to
Eq. \eqref{eq:app_PQ_cond_out}.

To study the behaviour in the critical region $\sqrt{V|\phi_0''|} t^* = O(1)$ as $V\gg 1$, it is 
convenient to notice that the steepest descent  path 
$\Gamma_a + \Gamma_b$ 
can be deformed into the imaginary axis $x=0$ without changing the value of the integral.
Taking $t= iy/\sqrt{V|\phi_0''|}$ the integral becomes:
\begin{align}
\label{eq:NM_D}
    \norm_{-}(\xi) &= \frac{1}{[V|\phi_0''|]^{1/4}} 
                       \int_{-\infty}^{+\infty}   \frac{dy}{2\pi}\, (i y)^{-1/2} e^{-y^2/2 - i\xi y}
                       \nonumber \\
                       &=
                       \frac{1}{\sqrt{2\pi}} \frac{e^{-\xi^2/4}\, D_{-1/2}(\xi)}{[V|\phi_0''|]^{1/4}}\, ,
\end{align}
where $\xi$ is defined in Eq. \eqref{eq:app_xi}, 
which with Eq. \eqref{eq:app_PQ_M} leads to Eq. \eqref{eq:app_PQ_in_norm}. 
However, notice that now $\xi < 0$ because we are in the condensed phase.
Using the asymptotic expansion $D_{-1/2}(-|\xi|) \sim e^{\xi^2/4}\, / \sqrt{|\xi|}$ valid for
$-\xi\gg 1$ Eq. \eqref{eq:app_PQ_in_norm} reduces to
\be
\label{eq:app_PQ_match_cond}
  P_{\rm match}(Q,V) =  \frac{e^{-V \phi(-\lambda_0^{-};q) }}
             {\sqrt{2\pi V (q_\delta^{-} - q)}},
\ee
and matches asymptotically Eq. \eqref{eq:app_PQ_cond_out}.
Thus, from  Eq. \eqref{eq:app_PQ_unif}, the uniform asymptotic approximation   to 
Eq. \eqref{eq:app_PQ} as $V \gg 1$ for $q \leq q_\delta^{-}$
reads
\be
\label{eq:app_PQ_cond_unif}
  P(Q;V) = {\cal N}_{-}(\xi) \,  e^{-V \phi(-\lambda_0^{-};q)}.
\ee

Notice that close to $q_\delta^{-}$, inside the critical region,  $P(Q;V)$ is 
given by Eq. \eqref{eq:app_PQ_in_norm} regardless of $q$ being larger or smaller than
$q_\delta^{-}$. The only difference is the sign of $\xi$. As a consequence
$P(Q;V)$ is regular across the transition between the two phases. The singularity shows up only
in the strict $V\to\infty$ limit.

\section{}
\label{app:alpha0}

In this Appendix we shortly discuss the  case of $\alpha = 0$, in which case the frequency 
$\omega(\mathbf{k})$   becomes $\mathbf{k}$-independent and the
condensation transition disappears.
If $\alpha=0$ the system reduces to that of $N$ harmonic oscillators of equal frequency 
$\omega$ and the Eq. \eqref{dis.6} is replaced by:
\be
P(Q,N) = \int_{\Gamma}\frac{d z}{2\pi i} \, e^{-N\phi(z;q)},
\label{eq:app_dis.6}
\ee
where 
\be
\phi(z;q) = q z + \frac{1}{2}
     \ln \left[ \frac{(\lambda_{+}- z)(z + \lambda_{-})}{\lambda_{+} \lambda_{-}} \right],
\label{eq:app_dis.4bis}
\ee
and $q=Q/N$ is the exchanged heat per oscillator.
In the large $N$ limit the integral is dominated by the saddle point $x^*$ on the real axis 
located at the stationary point of $\phi(z;q)$:
\be
  \phi'(x^*; q) = 0 \ \Rightarrow\ 
      q = \frac{1}{2}\left[\frac{1}{\lambda_+ - x^*}- \frac{1}{\lambda_- + x^*}\right].
\ee
Solving this equation one finds
\be
\label{eq:app_xstar}
  x^* = \frac{1}{2}(\lambda_+ - \lambda_-) + \frac{1}{2q}\bigl(\sqrt{\Delta} - 1\bigr),
\ee
with $\Delta = 1 + q^2(\lambda_+ + \lambda_-)^2$. It is not difficult to see that
$-\lambda_- < x^* < \lambda_+$ for all value of $q$ and hence
condensation cannot occur.

Near the saddle point the steepest descent path is parallel to the imaginary axis, and
\be
\label{eq:app_PQ_norm}
  P(Q,N) = \frac{1}{\sqrt{-2\pi N \phi''(x^*;q)}}\, e^{-N \phi(x^*;q)}.
\ee
as $N\gg1$. 
Using Eq. \eqref{eq:app_xstar} one finds
\be
  \phi(x^*;q) = \frac{1}{2} \left[ (\lambda_+ - \lambda_-) q +\sqrt{\Delta} - 1 + 
                                                \ln\Bigl[\frac{a}{2q^2}(\sqrt{\Delta}-1)\Bigr]
                                           \right],
\ee
and
\be
  \phi''(x^*;q) = - \frac{q^2(\lambda_+ + \lambda_-)^2 + (\sqrt{\Delta} - 1)^2}{(\sqrt{\Delta} - 1)^2}\,q^2.
\ee
Hence
\be
 \frac{1}{N}{\mathcal K}(Q) = (\lambda_- - \lambda_+)\, q
\ee
because $\Delta$ is even in $q$.


\begin{thebibliography}{99}
\bibitem{nota} To perform the asymptotic analysis of Eq. \eqref{dis.6}
  we assume that the variable $z$ and all parameters entering into the
  integral have been made dimensionless.  The correct dimension of
  $P(Q,V)$ can be reestablished by multiplying the result by, e.g,
  $\lambda_0^{-}$, or any other parameter with dimension $[Q]^{-1}$.
  This factor, nevertheless, can be always absorbed into the
  normalization term and is not considered in the computation.
\end{thebibliography}
\end{document}